\documentclass[letterpaper,twocolumn,showpacs,prl,aps]{revtex4-1}

\usepackage{amsmath}  
\usepackage{amsfonts} 
\usepackage{graphicx} 
\usepackage{amssymb}

\begin{document}

\title{Anomalous screening in two-dimensional materials with an extremum ring in the dispersion law}

\author{Eugene B. Kolomeisky$^{1}$ and Joseph P. Straley$^{2}$}

\affiliation
{$^{1}$Department of Physics, University of Virginia, P. O. Box 400714,
Charlottesville, Virginia 22904-4714, USA\\
$^{2}$Department of Physics and Astronomy, University of Kentucky,
Lexington, Kentucky 40506-0055, USA}

\date{\today}

\begin{abstract}
A variety of two-dimensional materials possess a band structure with an energy extremal ridge
along a ring in momentum space. Examples are biased bilayer graphene, and surfaces and interfaces
with a Rashba spin-orbit interaction where at low doping the carriers fill an annulus. This topological
feature causes an anomalous screening behavior, which we study using the Thomas-Fermi theory.
Specifically, reducing the doping is predicted to enhance the linear screening response, while at zero
doping the size of the screening cloud surrounding a Coulomb impurity is found to increase as the cube root of the impurity charge.
\end{abstract}


\maketitle

Screening is an important manifestation of the electron-electron interaction.  Linear screening is usually associated with the presence of an equilibrium density of free charge carriers which can redistribute themselves in response to weak external disturbances, thus reducing their effect.  Screening ceases to be linear when the perturbation is strong or the charge carriers are absent.  The efficiency of linear screening can be quantified by the Debye screening length, which may be viewed as a penetration depth for the electric field inside the material.  For instance, the electromagnetic shielding property of good conductors is due to the screening length being of atomic scale.  It is expected that having more charge carriers improves screening, so that the screening length is a decreasing function of carrier concentration which is indeed the case \cite{AM}.  

This paradigm of screening is based on three-dimensional experience, but is partly challenged in two dimensions:  the electron gases found in some semiconductor heterojunctions have a screening length which is independent of the carrier density \cite{AFS},  while undoped and unbiased bilayer graphene exhibits linear screening with a well-defined screening length \cite{KSA}.  

The goal of this paper is to show that when the dispersion law of quasiparticles has a degenerate ring of energy extrema and the occupied states inhabit an annulus, the dependence of the screening length on the carrier density is anomalous:  decreasing the doping \textit{decreases} the screening length.  This behavior is a consequence of the pseudo-one-dimensional character of the zero-point motion in the system brought about by the non-trivial topology of the occupied states in momentum space.   This conclusion affects a large class of laboratory two-dimensional systems, specifically surfaces, interfaces and heterojunctions with Rashba spin-orbit interaction \cite{Rashba_materials} and a variety of few-layer systems \cite{few_layer_materials};  a notable representative of the latter group is biased bilayer graphene  \cite{bilayer}.  The only other example of anomalous screening known to us is that of a three-dimensional electron gas in a very strong magnetic field \cite{SE}.  We additionally argue that despite the anomalous behavior of the screening length, materials with an extremum ring in the band structure will undergo a transition into a Mott insulating state \cite{Mott} upon decrease of the carrier density.   The effect of anomalous screening also manifests itself in the non-linear regime, in the properties of the screening cloud surrounding Coulomb impurity. We find a qualitatively stronger screening response than that in the standard theory of screening in three dimensions.  This is unusual because it is expected on general grounds that screening due to charges that are free to move in all three dimensions should be more efficient than in the case where the charges are confined to a plane in three-dimensional space.  We also discuss manifestations of the three-dimensional version of the effect and point out and explore its mapping onto the problem of an atom in a strong magnetic field \cite{Kadomtsev}.  

There is a number of experimentally verifiable implications of our results.  Here bilayer graphene seems to be the perfect system to study.   The fact that its band structure can be tuned between quadratic and annular through the electric field effect \cite{bilayer} makes it the ideal testbed for some of the consequences, for instance, those regarding the Coulomb impurity problem.

Without loss of generality we assume that the quasiparticles are the electrons that (in the regime of interest) obey the Bychkov-Rashba (BR) dispersion law  \cite{BR}      
\begin{equation}
\label{dispersion_law}
\varepsilon(\textbf{k})=\frac{\hbar^{2}}{2m}\left (k-k_{0}\right )^{2}
\end{equation}
where $\textbf{k}$ is the two-dimensional wavevector, $k=|\textbf{k}|$, and $m$ is the electron effective mass.   The dispersion law has a degenerate minimum along a circle of radius $k_{0}$ which is somewhat tunable \cite{Rashba_materials,few_layer_materials,bilayer}.  The level of doping will be characterized by the chemical potential $\mu$ whose zero is chosen at $k=k_{0}$;  the electrons are supplied by uniformly distributed donors.   In the range of doping we are interested in, $0\leqslant \mu \leqslant\hbar^{2}k_{0}^{2}/2m$, all the momentum states sandwiched between circles of inner radius $k_{1}=k_{0}-\sqrt{2m\mu/\hbar^{2}}$ and outer radius $k_{2}=k_{0}+\sqrt{2m\mu/\hbar^{2}}$ are occupied, and higher energy BR bands \cite{BR} play no role.  While the dispersion law (\ref{dispersion_law}) adequately describes the Rashba materials \cite{Rashba_materials} within the stated range of doping, for few-layer substances \cite{few_layer_materials} its range of applicability is narrowed to the vicinity of its minimum $k=k_{0}$.  Moreover, for the BR electrons the spectral degeneracy is lifted by the spin-orbit interaction which may not be the case for few-layer materials \cite{few_layer_materials} where spin and/or valley degeneracies may remain.  In what follows the latter possibility will be ignored in the interest of simplicity; only a simple modification of numerical factors would be needed for applications to a specific few-layer system.

The equilibrium density $n(\mu)$ and the chemical potential $\mu(n)$ of a non-interacting BR electron gas are given by
\begin{equation}
\label{n_of_mu}
n(\mu)=\int_{k_{1}}^{k_{2}}\frac{2\pi k dk}{(2\pi)^{2}}=\frac{k_{0}}{\pi}\sqrt{\frac{2m\mu}{\hbar^{2}}},~~~ \mu(n)=\frac{\pi^{2}\hbar^{2}n^{2}}{2mk_{0}^{2}}
\end{equation}
Even though the underlying electron system is two-dimensional, the quadratic dependence of the chemical potential $\mu$ on the particle density $n$ is a signature of a one-dimensional Fermi gas.  This is a consequence of the circle of minima in momentum space ({\ref{dispersion_law}) and the underlying reason behind the effect of anomalous screening. 

The macroscopic response of the BR system of interacting electrons to the presence of external disturbances  can be studied using the Thomas-Fermi (TF) method \cite{LL3} which is known to be reliable in the long-wavelength limit;  it is also applicable in the regime where the screening is non-linear \cite{AM}.  In order to provide a broader context for comparison of our results with what is known, we begin by outlining long-wavelength screening properties of a generic two-dimensional electron gas.  

The central object of the TF theory is the total potential $\varphi(\textbf{r})$ felt by an electron at a two-dimensional position $\textbf{r}$ which is due to the external potential $\varphi_{ext}(\textbf{r})$ and 
to the potential caused by the net local charge due to other electrons of density $n(\textbf{r})$ and
donors of density $n_{0}$:
 \begin{equation}
\label{scpotential}
\varphi(\textbf{r})=\varphi_{ext}(\textbf{r})-\frac{e}{\kappa}\int \frac{n(\textbf{r}' )- n_{0}}{|\textbf{r}-\textbf{r}'|}d^{2}r'
\end{equation}
where $\kappa$ is the background dielectric constant.  The integral is over the surface where the charge resides, but the Coulomb interaction has three-dimensional form since the fields extend into space. 

The TF approximation is that $e\varphi$ mimics a local change in chemical potential 
\begin{equation}
\label{equilibrium}
e\varphi(\textbf{r})=\mu[n(\textbf{r})]-\mu(n_{0})
\end{equation}

In the linearized theory of screening an approximation $n(\textbf{r})-n_{0}=(e\partial n_{0}/\partial \mu)\varphi$ is made followed by  the Fourier transformation of Eq.(\ref{scpotential}).   The outcome is an expression for the wave vector dependent static dielectric function of the two-dimensional electron gas \cite{AFS}
\begin{equation}
\label{dielectric-constant}
\epsilon(\textbf{k})=\kappa \left (1+\frac{q_{s}}{k}\right ), ~~~q_{s}=\frac{2 \pi e^{2}}{\kappa} \frac{\partial n_{0}}{\partial \mu}
\end{equation}
where $q_{s}^{-1}$ is the Debye screening length whose doping dependence is captured by the inverse density of states $\partial \mu/\partial n_{0}$.  Several previously studied cases are now worth examining: 

(i)  monolayer graphene:  the band structure exhibits the Dirac dispersion law ($\varepsilon \propto k$). As the doping is reduced, the density of states decreases, leading to a screening length that diverges $q_{s}^{-1}\propto \mu^{-1}$ \cite{graphene_review}.  From the viewpoint of three-dimensional experience this may be classified as "normal" screening.     

(ii)  electron gases with a parabolic dispersion law $\varepsilon \propto k^{2}$ (including unbiased bilayer graphene):  the chemical potential is proportional to the doping with the result that the density of states and screening length are independent of the doping \cite{AFS}.  This may be viewed as an example of "marginal" screening. 

(iii)  For the BR electron gas, employing the expression for the chemical potential (\ref{n_of_mu}), we find
\begin{equation}
\label{supersceening}
q_{s}^{-1}=\frac{\pi b}{2k_{0}^{2}}n_{0}=\frac{b}{2}\sqrt{\frac{2m\mu}{\hbar^{2}k_{0}^{2}}},~~~~~~~~~ b=\frac{\kappa \hbar^{2}}{me^{2}},
\end{equation}
where $b$ is the Bohr radius for the material.  Reduction in doping decreases the screening length to ever smaller values,which will dramatically increase the screening response.  This anomalous screening response is a consequence of the pseudo-one-dimensional form of the density of states $\partial n_{0}/\partial \mu$ which in turn is due to the ring of minima in the dispersion law (\ref{dispersion_law}).  For the same reason a BR quasiparticle exhibits unusual binding properties in short-range and Coulomb potentials \cite{Chaplik,Skinner}.  This linearized theory of screening is applicable for slowly varying external potentials satisfying the condition $k\ll q_{s}$.

The screening length cannot become arbitrarily small, however.  For sufficiently low doping the assumption that the electrons are free to move inevitably breaks down.  In our theory this is hidden within the approximation of a uniform distribution of donors by the neutralizing charge background of density $n_{0}$.  The free-electron assumption fails when the overlap of the wave functions of the electrons that could be bound to neighboring donors gets below a certain threshold value.  This is expected to happen at a critical density $n_{c}$ given by the two-dimensional version of the Mott criterion \cite{Mott}:
\begin{equation}
\label{Mott_transition}
n_{c}a^{2}\simeq 1,~~~~~a=\frac{b}{2\ln(k_{0}b)}, ~~\ln(k_{0}b)\gg1
\end{equation}
where $a$ is the localization length of the BR electron bound to a singly-charged donor \cite{Skinner}.  The energy scale corresponding to the threshold concentration (\ref{Mott_transition}) is $\mu_{c}\simeq (\hbar^{2}k_{0}^{2}/m)(k_{0}a)^{-4}$, which is within the energy range where the topology of occupied states is annular.  Our theory applies to the $n>n_{c}$ regime where conductivity is metallic.  For $n<n_{c}$ and low temperature the mechanism of conductivity is thermal activation. At the Mott threshold (\ref{Mott_transition}) the screening length ($\ref{supersceening}$) $q_{s}^{-1}\simeq b (\ln(k_{0}b)/k_{0}b)^{2}$ has atomic scale.  However, it is possible that for very short screening lengths the Mott transition may be replaced by a transition into some broken symmetry phase, as discussed in the literature \cite{Castro, Stauber}. 
  
The linear theory of screening generally breaks down at zero doping ($\mu(n_{0}=0)=0$), and then $n(e\varphi)$ implied by Eq.(\ref{equilibrium}) has to be substituted into Eq.(\ref{scpotential}).  The outcome is an integral equation encompassing various previously studied systems, as follows:  

(i)  When $n\propto (e\varphi)^{2}$ (the Dirac dispersion law), one recovers the non-linear TF theory of screening in undoped monolayer graphene \cite{graphene_TF}.

(ii) For undoped bilayer graphene (which has a parabolic dispersion law) one finds $n\propto e\varphi$ which  gives an "accidentally" linear theory of screening \cite{KSA}; this system is marginal.

(iii) For the BR electron system, Eq. (2) implies a relationship between the electron density and potential
\begin{equation}
\label{n_of_phi}
n=\frac{k_{0}}{\pi}\sqrt{\frac{2me\varphi}{\hbar^{2}}}   .
\end{equation}
This leads to another non-linear TF equation          
\begin{equation}
\label{non-linear_TF_equation}
\varphi(\textbf{r})=\varphi_{ext}(\textbf{r})-\frac{k_{0}}{\pi}\sqrt{\frac{2e}{\kappa b}}\int \frac{\sqrt{\varphi(\textbf{r}')}d^{2}r'}{|\textbf{r}-\textbf{r}'|}
\end{equation}
The same equation can be obtained by minimization of the TF energy functional
\begin{eqnarray}
\label{2d_pseudo_magnetic_TF_functional}
E[n(\textbf{r})]&=&\frac{\pi^{2}\hbar^{2}}{6mk_{0}^{2}}\int n^{3}(\textbf{r})d^{2}r
-e\int \varphi_{ext}(\textbf{r})n(\textbf{r})d^{2}r\nonumber\\
&+&\frac{e^{2}}{2\kappa}\int\frac{n(\textbf{r})n(\textbf{r}')d^{2}rd^{2}r'}{|\textbf{r}-\textbf{r}'|}
\end{eqnarray}  
combined with the definition of the total potential (\ref{scpotential}) and Eq.(\ref{n_of_phi}).  The first term of the functional (\ref{2d_pseudo_magnetic_TF_functional}) is the kinetic energy of the filled states, $T$, while the second and the third are the interaction energy of the electrons with the source, $U_{es}$, and the energy of the electron-electron-interaction, $U_{ee}$.   Several useful exact relationships (virial theorems) can be established between these contributions when the external potential is that of a point charge ($\varphi_{ext}=Ze/\kappa r$) by employing the extremal property of the functional (\ref{2d_pseudo_magnetic_TF_functional}).  Specifically, if $n^{*}(r)$ is the density minimizing the functional, we can consider the effect on $E$ of an small rescaling of the amplitude, $n^{*}\rightarrow (1+\gamma)n^{*}(r)$, and position, $n^{*}\rightarrow n^{*}[(1+\gamma)r]$, and then imposing the condition of extremum $(\partial E/\partial \gamma)_{\gamma=0}=0$.  This generates the identities $3T+U_{es}+2U_{ee}=0$ and $2T+U_{es}+3U_{ee}=0$.  One of their consequences is that the ground-state energy $E_{0}$ receives contributions in the ratio $T:U_{es}:U_{ee}=1:-5:1$.  Another consequence is that the ground-state energy can be computed from just one of the contributions into it, for example $E_{0}=T+U_{es}+U_{ee}=0.6U_{es}$.         

When the external potential is that of a point charge mimicking a Coulomb impurity of charge $Ze$, we seek a radially-symmetric solution to Eq.(\ref{non-linear_TF_equation}) in the form
\begin{equation}
\label{screening_function_definition}
\varphi(r)=\frac{Ze}{\kappa r}F\left (\frac{r}{\lambda}\right ), ~~~\lambda=\frac{1}{2}\left (\frac{bZ}{k_{0}^{2}}\right )^{1/3}
\end{equation}    
where $\lambda$ is the characteristic length scale.  The function $F(x)$ is subject to the boundary condition $F(0)=1$ and obeys the equation
\begin{equation}
\label{screening_function_equation}
F(x)=1-\frac{2}{\pi}x\int_{0}^{\infty}\sqrt{F(x')x'}dx'\frac{\textbf{K}\left (\frac{2\sqrt{xx'}}{x+x'}\right )}{x+x'}
\end{equation}
where $\textbf{K}(y)$ is the complete elliptic integral of the first kind.  Taking the $x\rightarrow \infty$ limit in (\ref{screening_function_equation}) we arrive at the identity $F(\infty)=1-\int_{0}^{\infty}\sqrt{F(x)x}dx$ which is internally consistent only if $F(\infty)=0$.  Indeed, the physically acceptable $F(\infty)$ is either zero (complete screening)  or a constant between zero and unity (a positively charged ion).  In the latter case, however,  the integral $\int_{0}^{\infty}\sqrt{F(x)x}dx$ diverges which is in contradiction with $1-F(\infty)$ being finite.   We conclude that the cloud of BR electrons completely screens external charge;  specifically $\int_{0}^{\infty}\sqrt{F(x)x}dx=1$ which means that solution to Eq.(\ref{screening_function_equation}) decreases faster than $1/x^{3}$ at large $x$.  Combining Eqs.(\ref{screening_function_definition}) and (\ref{n_of_phi}) we find an expression for the electron density
\begin{equation}
\label{electron_density}
n(r)= \frac{Z}{2\pi \lambda^{2}}\sqrt{\frac{\lambda}{r}F\left (\frac{r}{\lambda}\right )}
\end{equation}  
which shows that the density distribution around impurities of different $Z$ is similar with a characteristic length scale $\lambda \propto Z^{1/3}$.  The characteristic energy scale of the problem is $ Z^{2}e^{2}/\kappa \lambda \simeq  (e^{2}/\kappa)(k_{0}^{2}/b)^{1/3}Z^{5/3}$ which is also the estimate for the total ionization energy (teh negative of the ground-state energy).  These resemble the properties of an atom in a very strong magnetic field \cite{Kadomtsev}, and are very different from those for the regular TF atom where characteristic length scale decreases with $Z$ as $Z^{-1/3}$ and the total ionization energy behaves as $Z^{7/3}$ \cite{LL3}.  However, the atom in a strong magnetic field differs from the present situation in that the integration measures and position vectors are three-dimensional.     

The universal screening function $F(x)$ describing the electron cloud is a monotonically decreasing solution of the integral equation (\ref{screening_function_equation}).  In order to gain a better understanding of its properties it is useful to rewrite Eq.(\ref{screening_function_equation}) by employing Landen's transformation \cite{GR}
\begin{eqnarray}
\label{Landen}
F(x)=1&-&\frac{2}{\pi}\int_{0}^{x}\sqrt{F(x')x'}\textbf{K}\left (\frac{x'}{x}\right )dx'\nonumber\\
&-&\frac{2}{\pi}x\int_{x}^{\infty}\sqrt{\frac{F(x')}{x'}}\textbf{K}\left (\frac{x}{x'}\right )dx'
\end{eqnarray}
Let us assume for a moment that the electron cloud (like the TF atom in a strong magnetic field \cite{March}) has a boundary at $x=x_{0}$, i.e. $F(x\geqslant x_{0})=0$.  Evaluating both sides of Eq.(\ref{Landen}) at $x=x_{0}$ we then arrive at the identity $1=(2/\pi)\int_{0}^{x_{0}}\sqrt{F(x')x'}\textbf{K}(x'/x_{0})dx'$ which is consistent with the condition of complete screening $1=\int_{0}^{x_{0}}\sqrt{F(x')x'}dx'$ only if $x_{0}=\infty$.  We conclude that the screening cloud does not have a boundary and formally extends all the way to infinity, like its textbook counterpart \cite{LL3}. 

Eq.(\ref{Landen}) is also a convenient starting point for series expansion of $F(x)$ about the origin
\begin{equation}
\label{expansion}
F(x\rightarrow 0)=1+a_{2}x+a_{3}x^{3/2} +...
\end{equation}
where the first two expansion coefficients are given by
 \begin{equation}
\label{coefficients}
a_{2}=-\int_{0}^{\infty}\sqrt{\frac{F(y)}{y}}dy,a_{3}=2-\frac{2}{\pi}\int_{0}^{1}\sqrt{y}\textbf{K}(y)dy\approx1.48
\end{equation}     
and the rest can be expressed in terms of the slope $a_{2}=F'(0)$.  The latter has a useful physical interpretation that appears as the $r\rightarrow 0$ limit of the potential (\ref{screening_function_definition}) is taken:  $\varphi(r\rightarrow 0)=Ze/\kappa r+a_{2}(Ze/\kappa \lambda)+...$ where the second term is the potential due to the electrons at the origin, $\varphi_{e}(0)$.  Then the interaction energy of the electrons with the source $U_{es}=Ze\varphi_{e}(0)$ (measured in $Z^{2}e^{2}/\kappa \lambda$ energy units adopted hereafter) \textit{is} the slope $a_{2}$.

We found a solution to the non-linear singular integral equation (\ref{screening_function_equation}) by making sequential modifications to $n(r)$, seeking the global minimum of the functional (\ref{2d_pseudo_magnetic_TF_functional}) and inferring the screening function $F(x)$ via Eq.(\ref{electron_density}).  We subsequently verified that the resulting $F(x)$ satisfies the  integral equation (\ref{screening_function_equation}).  Our variational solver
determines the three terms of Eq. (\ref{2d_pseudo_magnetic_TF_functional}) separately, giving $T = 0.356$, $U_{es} = -1.802$, $U_{ee} = 0.367$, which are consistent with the virial theorems;  the value for $F'(0)$ is not evaluated very accurately but is consistent with the slope relationship $a_{2} = U_{es}$.  For the ground-state energy we find $E_{0}=- 1.079$.   The result for the screening function is shown in Figure 1.   Numerical analysis suggests that $F(x\rightarrow \infty)\propto x^{-5}$.  This is a qualitatively stronger screening response than that in the standard TF atom where the screening function falls of as $1/x^{3}$ at $x$ large \cite{LL3}.  The large distance behavior of the screening function establishes the upper bound in the range of applicability of our theory ($b/Z\ll r\ll (b/k_{0}^{2})^{1/3} Z^{4/3}$) which is the scale beyond which there are only a few electrons present;  the lower bound is the same as that in the regular TF theory \cite{LL3}.          
\begin{figure}
\includegraphics[width=1.0\columnwidth, keepaspectratio]{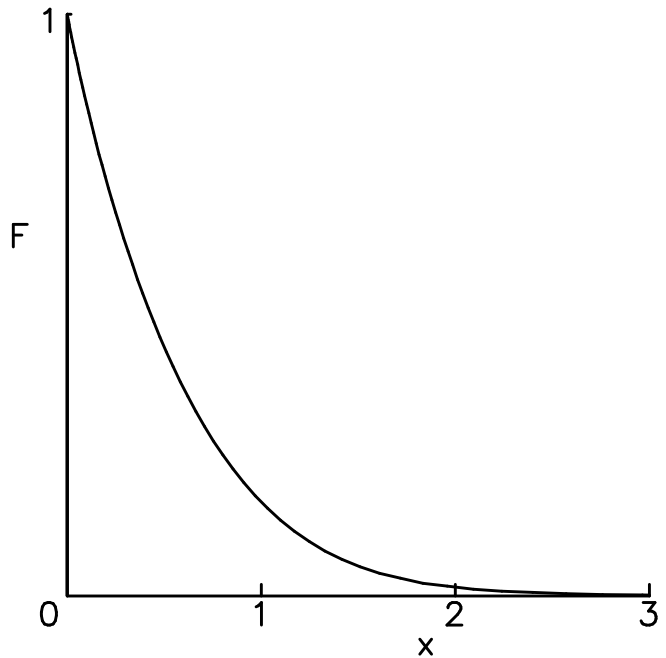} 
\caption{The screening function $F(x$), the solution to Eq.(\ref{screening_function_equation}).}
\end{figure}  
The range of applicability of our theory is then $Z\gg max [1, (k_{0}b)^{2/7}]$ which recovers the exact quantum-mechanical description in the $Z\rightarrow \infty$ limit  \cite{Lieb}.

The availability of materials combined with the ability to control $k_{0}$ makes the two-dimensional structures well-suited to test our predictions, but a similar anomalous screening can occur in three dimensions.  Generalizing Eq.(\ref{dispersion_law}) to a  three-dimensional dispersion law that is isotropic in momentum space, there would be a degenerate minimum along a spherical surface of radius $k_{0}$.  The roton minimum in the excitation spectrum of superfluid $He^{4}$ \cite{LL9} is an example.   Rotons are known to attract each other and form bound states for an arbitrarily weak two-body attraction \cite{Ruvalds}.  The one-dimensional character of this binding effect has been recognized \cite{Stephen}, thus lending support to the idea of anomalous screening in a Coulomb system.  It is worth mentioning that it may be possible to engineer roton-like dispersion laws in ultracold atomic gases as discussed in the literature \cite{Galitski,Demler}.     

Assuming the dispersion law (\ref{dispersion_law}) can be generated by a three-dimensional crystal structure (and leaving verification of the conjecture for future study), we proceed to conclusions.  When all the momentum states within a thin ($0\leqslant \mu \ll \hbar^{2}k_{0}^{2}/2m$) spherical layer are occupied, one would arrive at one-dimensional-like relationships $n\propto \mu^{1/2}$ and $\mu\propto n^{2}$, resembling Eqs.(\ref{n_of_mu}).  It is now straightforward to realize that what in two dimensions was referred to as an analogy with the problem of an electron gas in a strong magnetic field becomes a mapping in three dimensions.  Indeed, the three-dimensional Debye screening length $q_{s}^{-1} \propto(\partial\mu/\partial n_{0})^{1/2}$ \cite{AM} exhibits exactly the same anomalous $n_{0}^{1/2}$ behavior as that found for the three-dimensional electron gas in a strong magnetic field \cite{SE}.  Additionally, the electron cloud screening a $Z\gg1$ impurity is predicted to have an edge \cite{March};  the size of the cloud scales as $Z^{1/5}$ while the total ionization energy behaves as $Z^{9/5}$ \cite{Kadomtsev}. 
     
We are grateful to B. I. Shklovskii for a discussion which led to proper understanding of the condition of the Mott transition (\ref{Mott_transition}) and to E. I. Rashba for valuable comments.  This work was supported in part by US AFOSR Grant No. FA9550-11-1-0297.

\end{document}